\documentclass[prd,showpacs,showkeys,nofootinbib,twocolumn]{revtex4}

\usepackage{longtable}

\begin{document}

\title{Covariant equations of motion for test bodies in gravitational theories with general nonminimal coupling}

\author{Dirk Puetzfeld}
\email{dirk.puetzfeld@zarm.uni-bremen.de}
\homepage{http://puetzfeld.org}
\affiliation{ZARM, University of Bremen, Am Fallturm, 28359 Bremen, Germany} 

\author{Yuri N. Obukhov}
\email{yo@thp.uni-koeln.de}
\affiliation{Theoretical Physics Laboratory, Nuclear Safety Institute, Russian Academy of Sciences, B.Tulskaya 52, 115191 Moscow, Russia}

\date{ \today}

\begin{abstract}
We present a covariant derivation of the equations of motion for test bodies for a wide class of gravitational theories with nonminimal coupling, encompassing a general interaction via the complete set of 9 parity-even curvature invariants. The equations of motion for spinning test bodies in such theories are explicitly derived by means of Synge's expansion technique. Our findings generalize previous results in the literature and allow for a direct comparison to the general relativistic equations of motion of pole-dipole test bodies.
\end{abstract}


\maketitle


\section{Introduction}\label{introduction_sec}

In this work we derive the equations of motion for spinning test bodies in the context of gravitational theories with nonminimal coupling. Some of these theories have recently been investigated in \cite{Bertolami:etal:2007,Mohseni:2009,Mohseni:2010}. In particular, we generalize and extend the findings in \cite{Puetzfeld:Obukhov:2008:1} to the case in which the nonminimal coupling depends generally on the curvature of spacetime. Our results apply to a very large class of theories, i.e.\ the coupling is allowed to be a general function of the set of 9 parity-even curvature invariants. The covariant multipolar expansion method used to derive the equations of motion in the present work goes back to \cite{Synge:1960} -- it has also been utilized to derive the equations of motion of test bodies in Einstein's theory \cite{Dixon:1964}. Our findings therefore allow for a direct comparison of the new 'force' terms due to the nonminimal coupling procedure in a covariant fashion.
 
Without going into historical detail, we only mention that in the context of General Relativity multipolar methods of the kind employed in this work -- and variations of it -- were used in the works \cite{Mathisson:1937,Papapetrou:1951:3,Tulczyjew:1959,Tulczyjew:1962,Taub:1964,Dixon:1964,Madore:1969,Dixon:1970:1,Dixon:1970:2,Dixon:1979}. Similar methods have also been successfully applied in alternative gravity theories, see \cite{Stoeger:Yasskin:1979,Stoeger:Yasskin:1980,Nomura:Shirafuji:Hayashi:1991} and more recently in \cite{Puetzfeld:Obukhov:2007,Puetzfeld:Obukhov:2008}. A short timeline of works can be found in \cite{Puetzfeld:Obukhov:2007}. 

Note that the model under consideration does {\it not} belong to the very general class of gravitational models analyzed in \cite{Puetzfeld:Obukhov:2007} due to its nonminimal coupling prescription.  

The structure of the work is as follows: In section \ref{sec_models} we introduce two nonminimal coupling scenarios; in particular, we work out their respective conservation laws. With the conservation laws at hand, we derive the equations of motion for test bodies -- up to the pole-dipole order -- by means of a covariant multipolar method in section \ref{dipolar_eom_sec}. This is followed by our conclusion and outlook in \ref{conclusion_sec}. Appendix \ref{conventions_app} contains a brief overview of our conventions and notation, the dimensions of all quantities in the paper and a directory of used symbols can be found in tables \ref{tab_dimensions} and \ref{tab_symbols}. Appendix \ref{expansion_app} summarizes the expansion formulas used in our derivations.

\section{The models under consideration} \label{sec_models}

\subsection{Nonminimal $f(R)$ gravity} \label{subsec_model1}

In \cite{Bertolami:etal:2007} an extended version of a so-called $f(R)$ gravity theory was considered. Gravity theories in which the usual Einstein-Hilbert Lagrangian is replaced by an arbitrary function of the curvature scalar have attracted a lot of attention during the last few years see, e.g., the reviews \cite{Schmidt:2007,Straumann:2008,Nojiri:2011} and references therein. The $f(R)$ scenario was generalized even further in \cite{Bertolami:etal:2007} by the introduction of a nonminimal coupling term on the Lagrangian level. In particular, the following Lagrangian was put forward:
\begin{eqnarray}
L_{\rm tot} =  f_1\left(R\right) + \left[1+ \lambda f_2\left( R\right) \right] L_{\rm mat}. \label{ansatz_lagrangian}
\end{eqnarray} 
Here $f_1$ and $f_2$ are arbitrary functions of the curvature scalar $R$, and $L_{\rm mat}$ is the matter Lagrangian. The nonminimal coupling of matter and gravity is controlled by the constant $\lambda$. The general field equations -- in terms of the functions $f_1$ and $f_2$ and their derivatives -- are given in \cite{Bertolami:etal:2007}; their explicit form is irrelevant for the subsequent analysis though. 

In contrast to standard $f(R)$ gravity theories, the last term in (\ref{ansatz_lagrangian}) leads to a modification of the equations of motion. As was already shown in \cite{Koivisto:2006} (see also eq.\ (5) of 
\cite{Bertolami:etal:2007}) the usual conservation law -- as, for example, found in General Relativity -- is replaced by
\begin{eqnarray}
\nabla^i T_{ij} = \frac{\lambda f'_2}{1+\lambda f_2} \left( g_{ij} L_{\rm mat} - T_{ij} \right) \nabla^i R. \label{conservation}
\end{eqnarray}  
Here $f'_2\left(R\right):=df_2\left(R\right)/dR $ denotes a shortcut for derivatives of the unspecified function $f_2\left(R\right)$ of the curvature scalar and the energy-momentum tensor of matter is defined in the standard way by $\sqrt{-g}T_{ij} :=  -2\delta (\sqrt{-g}L_{\rm mat})/\delta g^{ij}$.

\subsection{General nonminimal gravity} \label{subsec_model2}

The above model can be generalized to
\begin{eqnarray}
L_{\rm tot} =  L_{\rm grav} + F L_{\rm mat}, \label{ansatz_lagrangian_model_2}
\end{eqnarray}
where both the gravitational field Lagrangian $L_{\rm grav} = L_{\rm grav}(g_{ij}, R_{ijk}{}^l)$ and the function $F = F(g_{ij}, R_{ijk}{}^l)$ can depend arbitrarily on the spacetime metric and the Riemannian curvature tensor. For example, in \cite{Mohseni:2009} both are assumed to be the functions of the Gauss-Bonnet scalar
\begin{eqnarray}
G=R^2-4 R_{ij} R^{ij} + R_{ijkl} R^{ijkl}. \label{definition_G}
\end{eqnarray}
This case belongs to the general class of models when the Lagrangian $L_{\rm grav} = L_{\rm grav}(i_1, i_2, i_3)$ and the function $F = F(i_1, i_2, i_3)$ both depend on the quadratic scalar invariants constructed from the components of the curvature tensor,
\begin{equation}\label{invariants}
i_1 = R^2,\qquad i_2 = R_{ij} R^{ij},\qquad i_3 = R_{ijkl} R^{ijkl}.
\end{equation}
As is well known \cite{Thomas:1934,Debever:1964,Carminati:1991}, a curved spacetime manifold of 4 dimensions is characterized by the 14 algebraically independent invariants constructed from the components of the Riemann tensor. There are two types of invariants: some of them are built of only the metric $g_{ij}$ and the curvature $R_{ijk}{}^l$, while others involve also the Levi-Civita totally antisymmetric tensor. The invariants of the first type are parity-even quantities (i.e., they do not change under space and time reflections), whereas the second type of invariants are parity-odd objects that change their sign under coordinate transformations which do not preserve orientation. There are 9 parity-even invariants and 5 parity-odd ones \cite{Debever:1964,Carminati:1991}. Here we will confine our attention to the general nonminimal coupling theories in which $F=F(i_1,\dots,i_9)$ is an arbitrary function of the set of the parity-even invariants that includes, in addition to the quadratic scalars (\ref{invariants}), the following cubic, quartic and quintic contractions: 
\begin{eqnarray}\label{inv4}
i_4 &=& R_{ij}{}^{kl}R_{kl}{}^{mn}R_{mn}{}^{ij},\\ 
i_5 &=& R^i{}_jR^j{}_kR^k{}_i,\quad
i_6 = R^i{}_jR^j{}_kR^k{}_lR^l{}_i,\label{inv56}\\
i_7 &=& R^{ij}D_{ij},\quad i_8 = D_{ij} D^{ij},\quad i_9 = D_{ij} D^{jk}R^i{}_k.\label{inv789}
\end{eqnarray}
Here we have denoted $D_{ij} := R_{iklj}R^{kl}$. The set (\ref{invariants})-(\ref{inv789}) is equivalent to the one reported in \cite{Debever:1964,Carminati:1991}, when the Riemann tensor is decomposed in terms of the Weyl and the traceless Ricci tensor. 

Generalized gravity theories with Lagrangians which are functions of the minimal independent set of curvature invariants have recently attracted some attention in the cosmological context, see \cite{Ishak:2009}, for example. 

The field equations for the model (\ref{ansatz_lagrangian_model_2}) are derived from the variation of the total action with respect to the spacetime metric. Denoting $\sqrt{-g}E_{ij} :=  2\delta (\sqrt{-g}L_{\rm grav})/\delta g^{ij}$, we find explicitly
\begin{eqnarray}
&&E^{ij} = FT^{ij} \nonumber\\
&&+ 2\sum^9_{A=1}\left[L_{\rm mat}F_AP_A^{ij} + \nabla_n\nabla_k \left(L_{\rm mat}F_A\pi_A^{ikjn}\right)\right].\label{graveqs}
\end{eqnarray}
Here $F_A = \partial F/\partial i_A$, $A=1,\dots,9$. The second line describes  the modification of the gravitational field equations due to the nonminimal coupling. Here, for the curvature quadratic invariants ($A=1,2,3$), 
\begin{eqnarray}
P_1^{ij} &=& - 2RR^{ij},\label{P1}\\
P_2^{ij} &=& - R^{ik}R^j{}_k - R^{iklj}R_{kl},\label{P2}\\
P_3^{ij} &=& - 2R^i{}_{klm}R^{jklm},\label{P3}
\end{eqnarray}
and
\begin{eqnarray}
\pi_1^{ikjn} &=& 2R(g^{in}g^{jk} - g^{kn}g^{ji}),\label{pi1}\\
\pi_2^{ikjn} &=& R^{in}g^{jk} - R^{kn}g^{ji} - R^{ij}g^{nk} + R^{kj}g^{ni},\label{pi2}\\
\pi_3^{ikjn} &=& 4R^{ikjn}.\label{pi3}
\end{eqnarray}
For the homogeneous cubic and quartic invariants $i_4, i_5, i_6$, given by (\ref{inv4})-(\ref{inv56}), we find
\begin{eqnarray}
P_4^{ij} &=& - 3R^{ikln}R_{ln}{}^{pq}R^j{}_{kpq},\label{P4}\\
P_5^{ij} &=& - {\frac 32}R^j{}_k{\stackrel {(2)}R}{}^{ik} -  {\frac 32}R^{iklj}{\stackrel {(2)}R}{}_{kl},\label{P5}\\
P_6^{ij} &=& - 2R^j{}_k{\stackrel {(3)}R}{}^{ik} - 2R^{iklj}{\stackrel {(3)}R}{}_{kl},\label{P6}
\end{eqnarray}
where we denoted ${\stackrel {(2)}R}{}^{ij} := R^{ik}R^j{}_k$ and ${\stackrel {(3)}R}{}^{ij} := R^{ik}R_{kl}R^{jl}$,  
\begin{eqnarray}
\pi_4^{ikjn} &=& 6R^{ik}{}_{pq}R^{jnpq},\label{pi4}\\
\pi_5^{ikjn} &=& {\frac 32}({\stackrel {(2)}R}{}^{in}g^{jk} - {\stackrel {(2)}R}{}{}^{kn}g^{ji} - {\stackrel {(2)}R}{}{}^{ij}g^{nk} + {\stackrel {(2)}R}{}^{kj}g^{ni}), \nonumber \\ 
\label{pi5}\\
\pi_6^{ikjn} &=& 2({\stackrel {(3)}R}{}^{in}g^{jk} - {\stackrel {(3)}R}{}^{kn}g^{ji} - {\stackrel {(3)}R}{}^{ij}g^{nk} + {\stackrel {(3)}R}{}^{kj}g^{ni}). \nonumber \\ 
\label{pi6}
\end{eqnarray}
Furthermore, for the mixed Riemann-Ricci quartic and quintic invariants (\ref{inv789}) we derive
\begin{eqnarray}
P_7^{ij} &=& - D^{ik}R^j{}_k - D^{jk}R^i{}_k - R^{iklj}D_{kl},\label{P7}\\
P_8^{ij} &=& - Y^{ik}R^j{}_k - Y^{jk}R^i{}_k - R^{iklj}Y_{kl}- {\stackrel {(2)}D}{}^{ij},\label{P8}\\
P_9^{ij} &=& - V^{ik}R^j{}_k - V^{jk}R^i{}_k - R^{iklj}V_{kl}- {\frac 12}R^{iklj}{\stackrel {(2)}D}{}_{kl}\nonumber\\ 
&-& {\frac 12}(R^{kl}D^i{}_kD^j{}_l + R^i{}_k{\stackrel {(2)}D}{}^{jk} + R^j{}_k{\stackrel {(2)}D}{}^{ik}),\label{P9}
\end{eqnarray}
where $Y^{ij} := R^{iklj}D_{kl}$ and ${\stackrel {(2)}D}{}^{ij} := D^{ik}D^j{}_k$, and 
\begin{eqnarray}
\pi_7^{ikjn} &=& R^{in}R^{jk} - R^{kn}R^{ji}\nonumber\\
&+& D^{in}g^{jk} - D^{kn}g^{ji} - D^{ij}g^{nk} + D^{kj}g^{ni},\label{pi7}\\
\pi_8^{ikjn} &=& D^{in}R^{jk} - D^{kn}R^{ji} - D^{ij}R^{nk} + D^{kj}R^{ni}\nonumber\\
&+& Y^{in}g^{jk} - Y^{kn}g^{ji} - Y^{ij}g^{nk} + Y^{kj}g^{ni},\label{pi8}\\
\pi_9^{ikjn} &=& U^{in}R^{jk} - U^{kn}R^{ji} - U^{ij}R^{nk} + U^{kj}R^{ni}\nonumber\\
&+& V^{in}g^{jk} - V^{kn}g^{ji} - V^{ij}g^{nk} + V^{kj}g^{ni}\nonumber\\
&+& {\frac 12}({\stackrel {(2)}D}{}^{in}g^{jk} - {\stackrel {(2)}D}{}^{kn}g^{ji} - {\stackrel {(2)}D}{}^{ij}g^{nk} + {\stackrel {(2)}D}{}^{kj}g^{ni}). \nonumber \\ 
\label{pi9}
\end{eqnarray}
Here $U^{ij} := R^{(i}{}_kD^{j)k}$ and $V^{ij} := R^{iklj}U_{kl}$.

One can verify the symmetry properties $P_A^{ij}=P_A^{ji}$, and
\begin{eqnarray}
\pi_A^{ikjn} = \pi_A^{[ik]jn} = \pi_A^{ik[jn]} = \pi_A^{jnik},\label{pisym}
\end{eqnarray}
for $A=1,\dots,9$. It is also straightforward to prove that the last term in (\ref{graveqs}) is symmetric in the indices $i,j$. 

The tensors (\ref{P1})-(\ref{pi3}) and (\ref{P4})-(\ref{pi9}) satisfy certain differential identities. The latter arise from the fact that the action-type integrals $I_A = \int d^4x\sqrt{-g}\,i_A$, $A=1,\dots,9,$ are invariant under general coordinate transformations. The corresponding Noether identities read
\begin{eqnarray}
\nabla_i\left({\frac 12}i_Ag^{ij} + P_A^{ij} + \nabla_n\nabla_k\pi_A^{ikjn}\right)= 0.\label{N0}
\end{eqnarray}
In view of the skew symmetry (\ref{pisym}), $\nabla_i\nabla_n\nabla_k \pi_A^{ikjn} = \nabla_{[i}\nabla_{n]}\nabla_k\pi_A^{ikjn}$. Then, using the definition of the curvature (\ref{curvature_def}), we rewrite the last term as 
\begin{eqnarray}\label{ddd}
\nabla_i\nabla_n\nabla_k\pi_A^{ikjn} = {\frac 12}R_{ikl}{}^j\nabla_n\pi_A^{ikln}.
\end{eqnarray}
As a result, the Noether identities (\ref{N0}) are recast into
\begin{eqnarray}
\nabla_iP_A^{ij} + {\frac 12}R_{ikl}{}^j\nabla_n\pi_A^{ikln} = -\,{\frac 12} \nabla^j\,i_A.\label{N1}
\end{eqnarray}
Actually, one can verify these differential identities directly by using the expressions (\ref{P1})-(\ref{pi3}) and (\ref{P4})-(\ref{pi9}) for $A=1,\dots,9$.

The nonminimal model (\ref{ansatz_lagrangian_model_2}) is invariant under the diffeomorphism (general coordinate) transformations. The corresponding Noether identities then tell us that $\nabla_iE^{ij} = 0$ identically, whereas on-shell (i.e., when the matter field equations are satisfied)
\begin{eqnarray}
&& \nabla_i\bigg\{FT^{ij} + 2\sum^9_{A=1}\left[L_{\rm mat}F_AP_A^{ij}\right. \nonumber\\ 
&& \left. + \nabla_n\nabla_k \left(L_{\rm mat}F_A\pi_A^{ikjn}\right)\right]\bigg\} = 0.\label{N2}
\end{eqnarray}
We can simplify this considerably by making use (\ref{ddd}) and the Noether identities (\ref{N1}). A direct check shows that (\ref{P1})-(\ref{pi3}) and (\ref{P4})-(\ref{pi9}) satisfy (for $A=1,\dots,9$)
\begin{eqnarray}
P_A^{ij} + {\frac 12}R_{kln}{}^j\pi_A^{klni} \equiv 0.\label{crucial}
\end{eqnarray}
Taking into account this crucial relation, the conservation law (\ref{N2}) is recast into the final form
\begin{eqnarray}\label{conservation2}
\nabla^i T_{ij} = {\frac 1F} \left(g_{ij} L_{\rm mat} - T_{ij} \right)\nabla^iF.
\end{eqnarray}  
This result generalizes the conservation law (\ref{conservation}) to the case in which $F = F(i_1,\dots,i_9)$ depends arbitrarily on the complete set of 9 parity-even curvature invariants (\ref{invariants})-(\ref{inv789}). Our derivation corrects the earlier studies \cite{Mohseni:2009,Mohseni:2010}. 

\section{Equations of motion}\label{dipolar_eom_sec}

In the following section, we derive the multipolar equations of motion for test bodies from the conservation law (\ref{conservation2}) by means of the covariant expansion technique from \cite{Synge:1960}. The multipolar moments extend the ones encountered in \cite{Dixon:1964} to the general nonminimal coupling scenario. 

The equations of motion will be explicitly worked out up to the dipole order; i.e., they are applicable to general spinning test bodies in theories with nonminimal coupling.

\subsection{Covariant moments \& multipolar expansion}

To begin with, we rewrite (\ref{conservation2}) as follows:
\begin{eqnarray}\label{conservation_expanded}
\nabla^i T_{ij} = \left( g_{ij} L_{\rm mat} - T_{ij} \right) \nabla^i A.
\end{eqnarray}  
Here we have introduced a scalar function $A\left(g_{ij}, R_{ijk}{}^l\right) := \log F$. In the following discussion, we are going to denote derivatives of this function simply by $A_i := \nabla_i A$, $A_{ij}:=\nabla_j \nabla_i A$, etc. Raising the indices and rewriting the covariant derivative in (\ref{conservation_expanded}) yields
\begin{eqnarray}
\nabla_i \widetilde{T}^{ij}&=& \left( \widetilde{\Xi}^{ij} - \widetilde{T}^{ij} \right) A_i. \label{conservation_rewritten}
\end{eqnarray}  
In the last equation, we introduced the quantity $\Xi^{ij}:=g^{ij} L_{\rm mat}$ as a shortcut. Densities of different quantities are denoted by a tilde ``$\widetilde{\phantom{A}}$''. 
 
We will now derive the equations of motion of a test body up to the dipole order by utilizing the covariant expansion method of Synge \cite{Synge:1960}. For this we need the following auxiliary formula for the absolute derivative of the integral of an arbitrary bitensor density $\widetilde{B}^{x_1 y_1}=\widetilde{B}^{x_1 y_1}(x,y)$:
\begin{eqnarray}
\frac{D}{ds} \int\limits_{\Sigma(s)} \widetilde{B}^{x_1 y_1} d \Sigma_{x_1} &=& \int\limits_{\Sigma(s)} \nabla_{x_1} \widetilde{B}^{x_1 y_1} w^{x_2} d \Sigma_{x_2} \nonumber \\
&& + \int\limits_{\Sigma(s)} v^{y_2} \nabla_{y_2} \widetilde{B}^{x_1 y_1} d \Sigma_{x_1}. \label{int_aux}
\end{eqnarray}
Here $v^{y_1}:=dx^{y_1}/ds$, $s$ is the proper time, and the integral is performed over a spatial hypersurface. Note that in our notation the point to which the index of a bitensor belongs can be directly read from the index itself; e.g., $y_{n}$ denote indices at the point $y$. Furthermore, we will now associate the point $y$ with the world-line of the test body under consideration. For additional comments regarding the explicit calculation of $w^a$ see the appendix of \cite{Dixon:1964}. Now we start by integrating (\ref{conservation_rewritten}):
\begin{eqnarray}
&&\!\!\!\!\! \int\limits_{\Sigma(s)} \sigma^{y_1}\!\!\cdots\sigma^{y_n} g^{y_0}{}_{x_0} \nabla_{x_1} \widetilde{T}^{x_0 x_1} w^{x_2} d \Sigma_{x_2} \nonumber \\
&=&\!\!\!\!\!  \int\limits_{\Sigma(s)}  \sigma^{y_1}\!\!\cdots \sigma^{y_n} g^{y_0}{}_{x_0} \!\left(\widetilde{\Xi}^{x_0 x_1} - \widetilde{T}^{x_0 x_1}\!\right)\! A_{x_1} w^{x_2} d\Sigma_{x_2}.  \label{start}
\end{eqnarray}
Here $\sigma$ denotes Synge's \cite{Synge:1960} world-function and $\sigma^y$ its first covariant derivative, cf.\ also appendix \ref{conventions_app} for a brief overview of our conventions. With the help of (\ref{int_aux}) we can rewrite the integral over the derivative of the energy-momentum tensor $\widetilde{T}^{x_0 x_1}$ as follows:
\begin{eqnarray}
&& \int\limits_{\Sigma(s)} \nabla_{x_1} \left(\sigma^{y_1}\cdots \sigma^{y_n} g^{y_0}{}_{x_0} \widetilde{T}^{x_0 x_1} \right) w^{x_2} d \Sigma_{x_2} \nonumber \\
&=& \frac{D}{ds} \int\limits_{\Sigma(s)} \sigma^{y_1} \cdots \sigma^{y_n}  g^{y_0}{}_{x_0} \widetilde{T}^{x_0 x_1} d \Sigma_{x_1} \nonumber \\
&-&  \int\limits_{\Sigma(s)} v^{y_{n+1}} \widetilde{T}^{x_0 x_1} \nabla_{y_3} \left(\sigma^{y_1} \cdots \sigma^{y_n} g^{y_0}{}_{x_0}  \right) d \Sigma_{x_1}. \label{inter_quad_1}
\end{eqnarray}
Equation (\ref{inter_quad_1}) allows us to rewrite the integral (\ref{start}), and thereby to derive the equations of motion at arbitrary order. 

We now introduce integrated moments \`{a} la Dixon in \cite{Dixon:1964}, i.e.
\begin{widetext}
\begin{eqnarray}
p^{y_1 \cdots y_n y_0}&:=& (-1)^n  \int\limits_{\Sigma(s)} \sigma^{y_1} \cdots \sigma^{y_n} g^{y_0}{}_{x_0}  \widetilde{T}^{x_0 x_1} d \Sigma_{x_1}, \label{p_moments_def} \\
t^{y_2 \cdots y_{n+1} y_0 y_1}&:=& (-1)^{n}  \int\limits_{\Sigma(s)} \sigma^{y_2} \cdots \sigma^{y_{n+1}} g^{y_0}{}_{x_0} g^{y_1}{}_{x_1}  \widetilde{T}^{x_0 x_1} w^{x_2} d \Sigma_{x_2}, \label{t_moments_def} \\
\xi^{y_2 \cdots y_{n+1} y_0 y_1}&:=& (-1)^{n}  \int\limits_{\Sigma(s)} \sigma^{y_2} \cdots \sigma^{y_{n+1}} g^{y_0}{}_{x_0} g^{y_1}{}_{x_1}  \widetilde{\Xi}^{x_0 x_1} w^{x_2} d \Sigma_{x_2}. \label{xi_moments_def}
\end{eqnarray}
Then the equation (\ref{start}) together with (\ref{inter_quad_1}), and the covariant expansions of the derivatives of the world-function and of the parallel propagator (see also appendix \ref{expansion_app}), yields 
\begin{eqnarray}
\frac{D}{ds} p^{y_1\dots y_ny_0} &=& t^{(y_1\dots y_n)y_0} - v^{(y_1} p^{y_2\dots y_n)y_0} -\,{\frac 12}R^{y_0}{}_{y'y'' y_{n+1}}\left(t^{y_1\dots y_ny_{n+1}y'y''} + p^{y_1\dots y_ny_{n+1}y'}v^{y''}\right)\nonumber\\
&& + \left(\xi^{y_1\dots y_ny'y_0} - t^{y_1\dots y_ny'y_0}\right)A_{y'} + \left( \xi^{y_1\dots y_ny_{n+1}y'y_0} - t^{y_1\dots y_ny_{n+1}y'y_0}\right)A_{y'y_{n+1}}\nonumber\\
&& + \sum\limits_{k=2}^\infty\,{\frac {1}{k!}}\biggl[\left(\xi^{y_1\dots y_ny_{n+1} \dots y_{n+k}y'y_0} - t^{y_1\dots y_ny_{n+1}\dots y_{n+k}y'y_0}\right)A_{y'y_{n+1}\dots y_{n+k}} \nonumber\\
&& +\,(-1)^k\gamma^{y_0}{}_{y'y''y_{n+1}\dots y_{n+k}}\left(t^{y_1\dots y_ny_{n+1}\dots y_{n+k}y'y''} + p^{y_1\dots y_n y_{n+1} \dots y_{n+k} y'}v^{y''}\right)\nonumber\\ \label{Dpn}
&& -\,(-1)^k\alpha^{(y_1}{}_{y'y_{n+1}\dots y_{n+k}}t^{y_2\dots y_n)y_{n+1}\dots y_{n+k}y_0y'} + (-1)^k\beta^{(y_1}{}_{y'y_{n+1}\dots y_{n+k}}p^{y_2\dots y_n)y_{n+1}\dots y_{n+k}y_0}v^{y'}\biggr].
\end{eqnarray}
\end{widetext}
In the dipole order, we keep only the multipole moments constructed up to the second order in the world-function $\sigma$. This truncates the infinite set of equations (\ref{Dpn}), with $n = 0,1,\dots,\infty$, to the three lowest-order equations. Namely, for $n = 2$ and $n=1$ we find
\begin{eqnarray}
v^{(y_1} p^{y_2)y_0} &=& t^{(y_1 y_2)y_0},\label{res_dipol_1}\\
\frac{D}{ds} p^{y_1 y_0} &=& t^{y_0 y_1} - v^{y_1} p^{y_0}\nonumber\\
&& +\,A_{y_2} \left(\xi^{y_1 y_0 y_2} - t^{y_1 y_0 y_2} \right).\label{res_dipol_2}
\end{eqnarray}
The constraint equation (\ref{res_dipol_1}) actually coincides with the one found in the general relativistic case. 

Finally, for $n=0$, we obtain
\begin{eqnarray}
\frac{D}{ds} p^{y_0} &=& -\,{\frac 12} R^{y_0}{}_{y_2 y_1 y_3} \left( v^{y_1} p^{y_3 y_2} + t^{y_3 y_2 y_1} \right)  \nonumber \\ 
&&+\,A_{y_1} \left(\xi^{y_0 y_1} - t^{y_0 y_1} \right)\nonumber \\ 
&&+\,A_{y_1y_2}\left(\xi^{y_2 y_0 y_1} - t^{y_2 y_0 y_1} \right).\label{res_dipol_3}
\end{eqnarray}

\subsection{Rewriting the equations of motion}\label{rewriting_eom_subsec}

Equations (\ref{res_dipol_1})-(\ref{res_dipol_3}) are all the information up to dipole order which we can extract from the integrated energy-momentum conservation law. The set of these three equations should be compared to the corresponding set of equations, which we derived in \cite{Puetzfeld:Obukhov:2008:1} in the context of Papapetrou's \cite{Papapetrou:1951:3} method. To make this even more explicit, we rewrite (\ref{res_dipol_1})-(\ref{res_dipol_3}) as follows:
\begin{eqnarray}
\frac{D}{ds} p^a &=& \frac{1}{2} R^a{}_{bcd} v^b s^{cd} + A_b \left(\xi^{ab} -t^{ab}\right) \nonumber \\
&& +\,A_{bc} \left(\xi^{cab} - t^{cab}\right), \label{dipl_eom_1}\\
\frac{D}{ds} s^{ab} &=& 2 v^{[b} p^{a]} + 2 A_c \left(t^{[ba]c} - \xi^{[ba]c} \right). \label{dipl_eom_2}
\end{eqnarray}
Here we have introduced the spin of the test body under consideration as $s^{ab} := 2 p^{[ab]}$, and switched back to the usual tensor notation, keeping in mind that all indices are now taken at the base-point $y$, which parametrizes the -- still completely arbitrary -- world-line.

The equations of motion for the momentum (\ref{dipl_eom_1}) and the spin (\ref{dipl_eom_2}) should be compared to our previous findings (\cite{Puetzfeld:Obukhov:2008:1},23) and (\cite{Puetzfeld:Obukhov:2008:1},29) in the context of a non-covariant multipole method. 

By utilizing the symmetries of the integrated $t$, $p$, and $\xi$ moments the dipole equations of motion can be further rewritten as
\begin{eqnarray}
\frac{D}{ds} {\cal P}^a &=& \frac{1}{2} R^a{}_{bcd} v^b{\cal S}^{cd} + \xi^{ab} \nabla_bF + \xi^{cab}\nabla_c\nabla_bF, \label{dipl_eom_1_1}\\ 
\frac{D}{ds}{\cal S}^{ab} &=& 2 v^{[b} {\cal P}^{a]} + 2\xi^{[ab]c}\nabla_cF. \label{dipl_eom_2_1}
\end{eqnarray}
Here we have introduced the generalized momentum and spin tensors as
\begin{eqnarray}
{\cal P}^a &=& Fp^a + p^{ba}\nabla_b F,\label{Pgen}\\
{\cal S}^{ab} &=& Fs^{ab}.\label{Sgen}
\end{eqnarray}
Note that in (\ref{dipl_eom_1_1}) and (\ref{dipl_eom_2_1}) the $t$ moments have been completely eliminated. 

A general interesting aspect of the present multipolar method is the fact, that the generalized momentum ${\cal P}^a$ follows from the equations of motion. To recall, within the context of Papapetrou's method we had to specify this quantity by hand to achieve the final form of the first equation of motion. In the present approach it is retrieved from the equation of motion for the spin, i.e.\ (\ref{dipl_eom_2_1}) yields:
\begin{eqnarray}\label{gen_mom_explicit}
{\cal P}^a = {\cal M} v^a + \dot{\cal S}^{ab} v_b - 2\xi^{[ab]c}v_b\nabla_cF, 
\end{eqnarray}
with ${\cal M}:={\cal P}^a v_a = Fm + p^{ab}v_b\nabla_aF$, where as usual, 
$m:=p^a v_a$.

It is worthwhile to mention the dual role played by the nonminimal function $F$: On the one hand, it ``rescales'' the ordinary momentum, spin and mass; and on the other hand, its gradients determine the force and torque that act on a particle in addition to the usual gravitational and Mathisson-Papapetrou forces. 

Finally, we note that -- as expected -- the motion of single-pole test bodies is also non-geodesic for the class of models under consideration. In this case, the geodesic equation, as encountered in General Relativity, is replaced by
\begin{eqnarray}
\frac{D}{ds}\left( Fmv^a\right) = \xi^{ab}\nabla_bF.\label{single_eom_1}
\end{eqnarray}
This can also be rewritten in an equivalent form
\begin{eqnarray}
m\dot{v}^a = \xi\left(\delta^a_b - v^av_b\right)\nabla^bA.\label{single_eom_2}
\end{eqnarray}
Here we recall that $\Xi^{ij}:=g^{ij} L_{\rm mat}$, and thus $\xi^{ab} = g^{ab}\xi$, with $\xi := \int\limits_{\Sigma(s)}L_{\rm mat}w^{x_2}d\Sigma_{x_2}$. As we see, a massive particle moves non-geodetically under the action of the ``pressure''-type force (\ref{single_eom_2}) produced by the nonminimal coupling function $F$.

\section{Conclusion}\label{conclusion_sec}

In \cite{Puetzfeld:Obukhov:2008:1} we employed Papapetrou's \cite{Papapetrou:1951:3} -- non-covariant -- approach to derive the equations of motion of the theory proposed in \cite{Bertolami:etal:2007}. The method utilized in the present work is more straightforward and has the benefit of being covariant. In 4 dimensions, there exist 14 algebraically independent curvature invariants \cite{Thomas:1934}. The results obtained in (\ref{dipl_eom_1_1}) and (\ref{dipl_eom_2_1}) generalize our previous findings to the case in which the nonminimal coupling depends arbitrarily on the Riemannian curvature of spacetime, with $F = F(i_1,\dots,i_9)$ being any function of the complete set of 9 parity-even curvature invariants (\ref{invariants})-(\ref{inv789}). The remarkably simple conservation law (\ref{conservation2}) derived in this work corrects the earlier erroneous results \cite{Mohseni:2009,Mohseni:2010}. Furthermore, our final equations of motion explicitly make clear, that the previous non-covariant method relies on a very specific transport process; i.e., the choice of the parallel propagator $g^{y_0}{}_{x_0}$. It is satisfying to see that we formally obtain the same equations of motion with the more general method, which still allows for a recovery of the previous results in a special case.

It is worthwhile to note that our results are compatible with the well-known general relativistic equations of motion for a spinning test body; i.e., in the minimal coupling case they reduce to the ones in \cite{Dixon:1964}, which in turn have also been derived by several authors by means of different multipolar approximation schemes \cite{Mathisson:1937,Papapetrou:1951:3,Tulczyjew:1959}.

\section*{Acknowledgements}
This work was supported by the Deutsche Forschungsgemeinschaft (DFG) through the grant LA-905/8-1 (D.P.). 

\appendix

\section{Conventions \& Symbols}\label{conventions_app}

\begin{table}
\caption{\label{tab_dimensions}Dimensions of the quantities.}
\begin{ruledtabular}
\begin{tabular}{cl}
Dimension (SI)&Symbol\\
\hline
&\\
\hline
\multicolumn{2}{l}{{Geometrical quantities}}\\
\hline
1 & $g_{ab}$, $\sqrt{-g}$, $\delta^a_b$, $g^{y_0}{}_{x_0}$\\
m& $s$, $dx^a$ \\
m$^2$ & $\sigma$\\
m$^{-1}$& $\Gamma_{ab}{}^c$ \\
m$^{-2}$& $R_{abc}{}^d$, $R_{ab}$, $R$ \\
&\\
\hline
\multicolumn{2}{l}{{Matter quantities}}\\
\hline
1& $v^a$  \\
kg/m$^2$ s & $T^{ab}$, $L$, $E^{ab}$, $\Xi^{ab}$\\
kg\,m/s & $p^a$, ${\cal P}^{a}$, $m$, ${\cal M}$, $\xi$ \\
kg\,m$^2$/s & $s^{ab}$, ${\cal S}^{ab}$\\
kg\,m$^{n+1}$/s & $t^{c_1 \dots c_n ab}$, $\xi^{c_1 \dots c_n ab}$, $p^{c_1 \dots c_n a}$ \\
&\\
\hline
\multicolumn{2}{l}{{Auxiliary quantities}}\\
\hline
1 & $F$, $I_A$, $A$, $\lambda f_2$ \\
m$^{-2}$ & $\pi_{1,2,3}^{abcd}$\\
m$^{-4}$ & $\pi_{4,5,7}^{abcd}$, $G$, $i_{1,2,3}$, $P^{ab}_{1,2,3}$, $D^{ij}$, ${\stackrel {(2)}R}{}^{ij}$\\
m$^{-6}$ & $\pi_{6,8}^{abcd}$, $P^{ab}_{4,5,7}$, $i_{4,5,7}$, $U^{ij}$, $Y^{ij}$, ${\stackrel {(3)}R}{}^{ij}$ \\
m$^{-8}$ & $\pi_{9}^{abcd}$, $P^{ab}_{6,8}$, $i_{6,8}$, $V^{ij}$,  ${\stackrel {(2)}D}{}^{ij}$ \\
m$^{-10}$ & $P^{ab}_{9}$, $i_{9}$ \\
m$^{4}$ & $F_{1,2,3}$\\
m$^{6}$ & $F_{4,5,7}$ \\
m$^{8}$ & $F_{6,8}$ \\
m$^{10}$ & $F_{9}$ \\
kg/m$^2$ s & $f_1$\\
m$^{-n+2}$&$\alpha^{y_0}{}_{y_1 \dots y_n}$, $\beta^{y_0}{}_{y_1 \dots y_n}$, $\gamma^{y_0}{}_{y_1 \dots y_n}$\\
&\\
\hline
\multicolumn{2}{l}{{Operators}}\\
\hline
m$^{-1}$& $\nabla_a$, $\partial_a$, $\frac{D}{ds} = $``$\dot{\phantom{a}}$'' \\
&\\
\end{tabular}
\end{ruledtabular}
\end{table}

\begin{table}
\caption{\label{tab_symbols}Directory of symbols.}
\begin{ruledtabular}
\begin{tabular}{ll}
Symbol & Explanation\\
\hline
&\\
\hline
\multicolumn{2}{l}{{Geometrical quantities}}\\
\hline
$g_{a b}$ & Metric\\
$\sqrt{-g}$ & Determinant of the metric \\
$\delta^a_b$ & Kronecker symbol \\
$x^{a}$, $s$ & Coordinates, proper time \\
$\Gamma_{a b}{}^c$ & Connection \\
$R_{a b c}{}^d$& Curvature \\
$\sigma$ & World-function\\
$g^{y_0}{}_{x_0}$ & Parallel propagator\\
$G$ & Gauss-Bonnet scalar\\
&\\
\hline
\multicolumn{2}{l}{{Matter quantities}}\\
\hline
$v^a$ & Velocity \\
$m$ & Mass \\
$p^a$ & Generalized momentum \\
$S^{ab}$ & Spin tensor \\
$T^{a b}$ & Energy-momentum tensor\\
$L$ & Lagrangian \\ 
$t^{c_1 \dots c_n ab}$, $\xi^{c_1 \dots c_n ab}$, $p^{c_1 \dots c_n a}$ & Integrated moments\\
&\\
\hline
\multicolumn{2}{l}{{Auxiliary quantities}}\\
\hline
$\lambda$ & Coupling constant \\
$f_1$, $f_2$, $F$ & Functions of the curvature\\
$i_A$ & Scalar curvature invariants\\
$F_A$ & Deriv.\ of $F$ w.r.t.\ to invariants\\
$P^{ab}_A$, $\pi_A^{abcd}$, $U^{ij}$, $V^{ij}$, $Y^{ij}$, & Shortcuts\\
$D^{ij}$, ${\stackrel {(2)}D}{}^{ij}$, ${\stackrel {(2)}R}{}^{ij}$, ${\stackrel {(3)}R}{}^{ij}$ &\\
$\alpha^{y_0}{}_{y_1 \dots y_n}$, $\beta^{y_0}{}_{y_1 \dots y_n}$, $\gamma^{y_0}{}_{y_1 \dots y_n}$& Expansion coefficients\\
&\\
\hline
\multicolumn{2}{l}{{Operators}}\\
\hline
$\partial_i$, $\nabla_i$ & (Partial, covariant) derivative \\ 
$\frac{D}{ds} = $``$\dot{\phantom{a}}$'' & Total derivative \\
``$[ \dots ]$''& Coincidence limit\\
&\\
\end{tabular}
\end{ruledtabular}
\end{table}

Our conventions for the Riemann curvature are as follows:
\begin{eqnarray}
&& 2 T^{c_1 \dots c_k}{}_{d_1 \dots d_l ; [ba] } \equiv 2 \nabla_{[a} \nabla_{b]} T^{c_1 \dots c_k}{}_{d_1 \dots d_l} \nonumber \\
& = & \sum^{k}_{i=1} R_{abe}{}^{c_i} T^{c_1 \dots e \dots c_k}{}_{d_1 \dots d_l} \nonumber \\
&& - \sum^{l}_{j=1} R_{abd_j}{}^{e} T^{c_1 \dots c_k}{}_{d_1 \dots e \dots d_l}. \label{curvature_def}
\end{eqnarray}
The Ricci tensor is introduced by $R_{ij} = R_{kij}{}^k$, and the curvature scalar is $R = g^{ij}R_{ij}$. The signature of the spacetime metric is assumed to be $(+1,-1,-1,-1)$.

In the following, we summarize some of the frequently used formulas in the context of the bitensor formalism (in particular for the world-function $\sigma(x,y)$), see, e.g., \cite{Synge:1960,DeWitt:Brehme:1960,Poisson:etal:2011} for the corresponding derivations. Note that our curvature conventions differ from those in \cite{Synge:1960,Poisson:etal:2011}. Indices attached to the world-function always denote covariant derivatives, at the given point, i.e.\ $\sigma_y:= \nabla_y \sigma$, hence we do not make explicit use of the semicolon in case of the world-function. We start by stating, without proof, the following useful rule for a bitensor $B$ with arbitrary indices at different points (here just denoted by dots):
\begin{eqnarray}
\left[B_{\dots} \right]_{;y} = \left[B_{\dots ; y} \right] + \left[B_{\dots ; x} \right]. \label{synges_rule}
\end{eqnarray}
Here a coincidence limit of a bitensor $B_{\dots}(x,y)$ is a tensor 
\begin{eqnarray}
\left[B_{\dots} \right] = \lim\limits_{x\rightarrow y}\,B_{\dots}(x,y),\label{coin}
\end{eqnarray}
determined at $y$. Furthermore, we collect the following useful identities: 
\begin{eqnarray}
&&\sigma_{y_0 y_1 x_0 y_2 x_1} = \sigma_{y_0 y_1 y_2 x_0 x_1} = \sigma_{x_0 x_1 y_0 y_1 y_2 }, \label{rule_1} \\
&&g^{x_1 x_2} \sigma_{x_1} \sigma_{x_2} = 2 \sigma = g^{y_1 y_2} \sigma_{y_1} \sigma_{y_2}, \label{rule_2}\\
&&\left[ \sigma \right]=0, \quad  \left[ \sigma_x \right] = \left[ \sigma_y \right]  = 0, \label{rule_3} \\
&& \left[ \sigma_{x_1 x_2} \right] =  \left[ \sigma_{y_1 y_2} \right] = g_{y_1 y_2}, \label{rule_4}\\ 
&& \left[ \sigma_{x_1 y_2} \right] =  \left[ \sigma_{y_1 x_2} \right] = - g_{y_1 y_2}, \label{rule_5}\\ 
&& \left[ \sigma_{x_1 x_2 x_3} \right] = \left[ \sigma_{x_1 x_2 y_3} \right] = \left[ \sigma_{x_1 y_2 y_3} \right] = \left[ \sigma_{y_1 y_2 y_3} \right] = 0, \nonumber \\ \label{rule_6}\\
&&\left[g^{x_0}{}_{y_1} \right] = \delta^{y_0}{}_{y_1}, \quad \left[g^{x_0}{}_{y_1 ; x_2} \right] = \left[g^{x_0}{}_{y_1 ; y_2} \right] = 0, \label{rule_7} \\
&& \left[g^{x_0}{}_{y_1 ; x_2 x_3} \right] = \frac{1}{2} R^{y_0}{}_{y_1 y_2 y_3}. \label{rule_8}
\end{eqnarray}

\section{Covariant expansions}\label{expansion_app}

Here we briefly summarize the covariant expansions of the second derivative of the world-function, and the derivative of the parallel propagator:
\begin{eqnarray}
\sigma^{y_0}{}_{x_1} &=& g^{y'}{}_{x_1}\biggl( -\,\delta^{y_0}{}_{y'}\nonumber\\
&& +\,\sum\limits_{k=2}^\infty\,{\frac {1}{k!}}\,\alpha^{y_0}{}_{y'y_2\!\dots \!y_{k+1}}\sigma^{y_2}\cdots\sigma^{y_{k+1}}\biggr)\!,\label{app_expansion_1}\\
\sigma^{y_0}{}_{y_1} &=& \delta^{y_0}{}_{y_1} \nonumber\\
&& -\,\sum\limits_{k=2}^\infty\,{\frac {1}{k!}}\,\beta^{y_0}{}_{y_1y_2\dots y_{k+1}} \sigma^{y_2}\!\cdots\!\sigma^{y_{k+1}}, \label{app_expansion_2} \\
g^{y_0}{}_{x_1 ; x_2} &=& g^{y'}{\!}_{x_1} g^{y''}{\!}_{x_2}\biggl({\frac 12} 
R^{y_0}{}_{y'y''y_3}\sigma^{y_3}\nonumber\\ 
&&\!+\!\sum\limits_{k=2}^\infty\,{\frac {1}{k!}}\,\gamma^{y_0}{}_{y'y''y_3\dots y_{k+2}}\sigma^{y_3}\!\cdots\!\sigma^{y_{k+2}}\!\biggr)\!,\label{app_expansion_3} \\
g^{y_0}{}_{x_1 ; y_2} &=& g^{y'}{\!}_{x_1} \biggl({\frac 12} R^{y_0}{}_{y'y_2y_3}\sigma^{y_3}\nonumber\\ 
&&\!+\!\sum\limits_{k=2}^\infty\,{\frac {1}{k!}}\,\gamma^{y_0}{}_{y'y_2y_3\dots y_{k+2}}\sigma^{y_3}\!\cdots\!\sigma^{y_{k+2}}\!\biggr).\label{app_expansion_4}
\end{eqnarray}
The coefficients $\alpha, \beta, \gamma$ in these expansions are polynomials constructed from the Riemann curvature tensor and its covariant derivatives. The first coefficients read as follows:
\begin{eqnarray}
\alpha^{y_0}{}_{y_1y_2y_3} &=& - \frac{1}{3} R^{y_0}{}_{(y_2y_3)y_1},\label{a1}\\
\beta^{y_0}{}_{y_1y_2y_3} &=& \frac{2}{3}R^{y_0}{}_{(y_2y_3)y_1},\label{be1}\\
\alpha^{y_0}{}_{y_1y_2y_3y_4} &=& - \frac{1}{2} \nabla_{(y_2}R^{y_0}{}_{y_3y_4)y_1},\label{al2}\\
\beta^{y_0}{}_{y_1y_2y_3y_4} &=& \frac{1}{2} \nabla_{(y_2} R^{y_0}{}_{y_3y_4)y_1},\label{be2}\\
\nonumber\\
\gamma^{y_0}{}_{y_1y_2y_3y_4}&=& \frac{1}{3} \nabla_{(y_3} R^{y_0}{}_{|y_1|y_4)y_2}.\label{ga}
\end{eqnarray}
In addition, we also need the covariant expansion of a usual vector:
\begin{eqnarray}
A_x = g^{y_0}{}_x\,\sum\limits_{k=0}^\infty\,{\frac {(-1)^k}{k!}} \, A_{y_0;y_1\dots y_k}\,\sigma^{y_1}\cdots\sigma^{y_k}.\label{Ax}
\end{eqnarray}

\bibliographystyle{unsrtnat}
\bibliography{covnonmin_bibliography}
\end{document}